\documentclass[useAMS,usenatbib,10pt]{report}
\usepackage{graphicx}

\title{The lack of carbon stars in the Galactic bulge }

\begin{document}
\begin{abstract}
In order to explain the lack of carbon stars in the Galactic bulge,
we have made a detailed study of thermal pulse - asymptotic giant
branch stars by using a population synthesis code. The effects of
the oxygen overabundance and the mass loss rate on the ratio of the
number of carbon stars to that of oxygen stars in the Galactic bulge
are discussed. We find that the oxygen overabundance which is about
twice as large as that in the solar neighbourhood (close to the
present observations) is insufficient to explain the rareness of
carbon stars in the bulge. We suggest that the large mass loss rate
may serve as a controlling factor in the ratio of the number of
carbon stars to that of oxygen stars.
\end{abstract}

\section{Introduction}
Stars with initial masses in a range of $0.9M_\odot<M<8M_\odot$ go
through the asymptotic giant branch (AGB) phase at the end of their
life. An AGB star has two nuclear burning shells: one is the burning
helium surrounded a carbon-oxygen core, and the other is the burning
hydrogen which is below a deep convective envelope. There exists an
intershell region between helium burning shell and hydrogen burning
shell. This region is rich in helium and carbon but has little neno
and oxygen.\cite{k02} There are two special features in AGB
evolution: thermal pulse (TP) and mass loss.\cite{v93,b91} In each
TP, the outer layers expand due to the high helium burning
luminosity. The strong expansion has an influence on extinguishing
the hydrogen shell. Then the deep convective envelope can penetrate
into the intershell region and mix with the products of internal
nucleosynthesis. This mixing event is named the third dredge up
(TDU). Following the dredge up, the hydrogen shell is re-ignited due
to the star contraction. The star enters a phase of quiescent
hydrogen burning known as the interpulse phase. In the current
understanding of the stellar evolution theory, the carbon stars
cannot be formed before the TP-AGB phase, the phase which occurs
after the first TP to the end of AGB. After each TP, the surface
abundances of TP-AGB stars may be modified by the TDU.\cite{i83}
Following the dredge up, a certain quantity of intershell matter is
brought up to the surface, and the surface abundance of $^{12}$C
becomes enriched. Under the repeated actions of TPs and TDUs, the
stars can become carbon stars with ratio $C/O>1$ in the envelopes if
they are efficient enough. In recent decades, many studies have
shown that the carbon stars in the Galactic bulge are very rare,
compared with the numbers in the solar neighbours.\cite{w93,t91} The
recent study\cite{f06} indicated that the stellar population of the
Galactic bulge does not contain any normal carbon stars except one
candidate. However, the rareness of the bulge carbon stars is still
a mystery. Feast\cite{f06} showed the lack of carbon stars in the
bulge could not be explained simply by an age effect or a high metal
abundance and suggested that the oxygen overabundance maybe modified
the ratio of the number of carbon stars to that of oxygen giants.

In this paper, using the synthetic models and the different
parameters of the initial oxygen abundance, we simulate TP-AGB stars
in the Galactic bulge and discuss the effect of oxygen overabundance
on the ratio of the number of carbon stars to that of oxygen TP-AGB
stars (O-rich stars). In addition, we attempt to find the other
controlling factors in the ratio of the number of carbon stars to
that of O-rich stars, for example, the mass loss. In Section 2, we
present our assumptions and describe some details of the algorithm.
In Section 3, we discuss the main results and the effects of the
different parameters. In Section 4, the main conclusions are given.

\section{The Model}
\subsection{Basic Parameters of the Monte Carlo Simulation }

For the  population synthesis of the single stars, the main input
parameters of the model are: (i) the initial mass function (IMF);
(ii) the lower and the upper mass cut-offs $M_{\rm l}$ and $M_{\rm
u}$ for the initial mass function; (iii) the relative age of the
single stellar population; (iv) the metallicity $Z$ of the
stars.\cite{h95,z02,h03}

Our primary mass distribution takes the IMF as\cite{ktg93}
\begin{equation}
\xi(m)dm=\left\{
\begin{array}{cc}
           0 & m\leq m_0,\\
           a_1m^{-1.3}dm & m_0<m\leq 0.5,\\
           a_2m^{-2.2}dm & 0.5<m\leq 1.0,\\
           a_2m^{-2.7}dm & 1.0<m<\infty,
\end{array}
 \right.
\end{equation}
where $\xi(m)dm$ is the probability with which a star has a mass (in
solar units) between $m$ and $m+dm$; $m_0=0.1M_\odot$;
$a_1=0.29056$; $a_2=0.15571$.

The lower and the upper mass cut-offs are $M_{\rm l}=0.1M_\odot$ and
$M_{\rm u}=120M_\odot$, respectively. If the initial mass is larger
than $4.0M_\odot$, the stars hardly become carbon stars because of
the hot bottom burning which turns $^{12}$C into $^{14}$N. At the
same time, the absence of the TDU prevents the carbon star from
forming when the initial mass is less than $1.5M_\odot$. Therefore,
we lay a strong emphasis on the stars with the initial masses
ranging from $1.5M_\odot$ to $4.0M_\odot$ in this paper.

The average value of [Fe/H] in the Galactic bulge is close to the
mean value for the solar neighborhood.\cite{a97} In our work, we
take the metallicity as $Z$=0.02 for convenience. The maximum age of
the single stellar population is 15 Gyr.

The recent work has shown that the star formation rate in the
Galactic bulge is $10-100M_\odot$ yr$^{-1}$.\cite{g01,z05} We assume
that the stars with total mass of about $50M_\odot$ are born in the
bulge every year.

\subsection{Synthesised TP-AGB Evolution}
The stellar evolution from the zero age main sequence to the first
thermal pulse is dealt with using the rapid evolution code by Hurley
et al.\cite{h00} After the first thermal pulse, we use a synthetic
model for TP-AGB. The changes of the chemical abundances on the
stellar surface during the giant branch (the first dredge up) and
the early AGB (E-AGB) phase (the second dredge up) can be
represented by the simple fitting formulae in Refs.\cite{i04,I04}.
In the following sections, the fitting formulae mainly come from
Refs.\cite{i04,I04}. They are comprised of a Levenberg-Marquart
gradient descent iterative $\chi^2$-minimization code.\cite{i04}

\subsubsection{The initial abundances}
It has been known that oxygen in the bulge M giants is overabundant
about twice larger than that in the solar neighborhood.\cite{ro05}
In our work, we take the initial abundances as those in
Ref.\cite{AG89} for $Z=0.02$ except oxygen abundance. To study the
effect of the oxygen overabundance on the ratio of the number of
carbon stars to that of O-rich stars, the oxygen overabundance is
taken as the corresponding value of coefficient $\theta$ in each
case involved in our model. The details are shown in Table 1. The
following data
are the initial abundances shown by the mass fractions in our model:\\
$^1$H=0.68720, \  \ \ \ \  \ \ \ \ \ \ \ \ \ \ \ \ $^4$He=0.29280,\\
$^{12}$C=2.92293$\times10^{-3}$, \  \ \ \ \  \ \ \ $^{13}$C=4.10800$\times10^{-5}$,\\
$^{14}$N=8.97864$\times10^{-4}$, \  \ \ \  \ \ \ \
$^{15}$N=4.14000$\times10^{-6}$,\\
$^{16}$O=8.15085$\times10^{-3}\times \theta$, \  \ \  $^{17}$O=3.87600$\times10^{-6}$,\\
$^{20}$Ne=2.29390 $\times10^{-3}$, \  \ \ \  \ \  $^{22}$Ne=1.45200$\times10^{-4}$.\\
The data for other elements are not shown here, and they do not
affect our results. The enrichment of $^{16}$O is equivalent to the
enhancements of $\alpha$ elements which change the stellar opacity.
Then, the enrichment of $^{16}$O may have some influence on the
stellar evolution. Salasnich et al\cite{sg00} have shown that the
enhancements of $\alpha$ elements have a very small influence on the
evolution of the low mass stars ($M<5M_\odot$). The carbon stars
originate from the stars with masses lower than 4$M_\odot$. In our
paper, we neglect the influence of the enrichment of $^{16}$O on the
stellar evolution. While the abundance of $^{16}$O increases, we
divide all metal element abundances by
$\frac{0.02+(\theta-1)^{16}{\rm O}}{0.02}$ in order to keep
$Z=0.02$, where $^{16}{\rm O}=8.15085\times10^{-3}$.
\subsubsection{Core mass during the first thermal pulse}
Using the rapid evolution code by Hurley et al,\cite{h00} we can
obtain the stellar mass ($M_{\rm 1TP}$) during the first thermal
pulse. The core mass during the first thermal pulse, $M_{\rm c,
1TP}$, is\cite{k02}
\begin{equation}
M_{\rm c, 1TP}=[-p_1(M_0-p_2)+p_3]f+(p_4M_0+p_5)(1-f),
\label{eq:mc1tp}
\end{equation}
where $f=(1+e^{(\frac{M_0-p_6}{p_7})})^{-1}$, and $M_0$ is the
initial mass in solar units. The coefficients $p_1$, $p_2$, $p_3$,
$p_4$, $p_5$, $p_6$ and $p_7$ are shown in Table 6 of
Ref.\cite{k02}.

\subsubsection{Luminosity, radius and interpulse period}
We use the prescriptions by Izzard et al.\cite{i04} The luminosity
is taken as the value calculated from Eq.(29) in Ref.\cite{i04}. We
define radius $R$ as $L=4\pi \sigma R^2T^4_{\rm eff}$, where
$\sigma$ is the Stefan-Boltzmann constant and $T_{\rm eff}$ is the
effective temperature of the star. The radius is taken as the value
calculated from Eq.(35) in Ref.\cite{i04}. The interpulse period
$\tau_{\rm ip}$ is
\begin{equation}
{\rm log}_{10}(\tau_{\rm ip}/\rm yr)=a_{28}(M_{\rm
c}/M_\odot-b_{28})-10^{c_{28}}-10^{d_{28}}+0.15\lambda^2,
\label{eq:tip}
\end{equation}
where the TDU efficiency $\lambda$ is defined in Section 2.2.4, and
the coefficients are\\
$a_{28}=-3.821,$\\
$b_{28}=1.8926,$\\
$c_{28}=-2.080-0.353Z+0.200(M_{\rm env}/M_\odot+\alpha-1.5),$\\
$d_{28}=-0.626-70.30(M_{\rm c,1TP}/M_\odot-\zeta)(\Delta M_{\rm
c}/M_\odot),$\\where $M_{\rm env}$ represents the envelope mass,
$\alpha$ is the mixing length parameter and equal to 1.75,
$\zeta={\rm log}(Z/0.02)$, and $\Delta M_{\rm c}$ is the change in
core mass defined as $\Delta M_{\rm c}=M_{\rm c}-M_{\rm c, 1TP}$.

\subsubsection{The minimum core mass for the TDU and the TDU efficiency}
\label{sec:lamb} The TDU can occur only in stars with masses above a
certain core mass $M^{\rm min}_{\rm c}$. Groenewegen and de Jong
took $M^{\rm min}_{\rm c}$ as a constant 0.58$M_\odot$.\cite{gj93}
Karakas et al\cite{k02} have found that $M^{\rm min}_{\rm c}$
depends on the stellar mass and the metallicity. They gave a fitting
formula
\begin{equation}
M^{\rm min}_{\rm c}=a_1+a_2M_0+a_3M_0^2+a_4M_0^3, \label{eq:mcmin}
\end{equation}
where coefficients $a_1$, $a_2$, $a_3$ and $a_4$ are shown in Table
7 of Ref.\cite{k02} and $M_0$ is the initial mass in solar units.
According to the carbon star luminosity function in the Magellanic
clouds, Marigo and Girardi\cite{mg07} have considered that the
predicted $M^{\rm min}_{\rm c}$ in Ref.\cite{k02} is high. In this
work, we take the value calculated from Eq.(\ref{eq:mcmin}) as
$M^{\rm min}_{\rm c}$.

It should be recalled that $M^{\rm min}_{\rm c}$=$M_{\rm c, 1TP}$ if
a stellar initial mass is $M_{\rm initial}\geq 4M_\odot$ or $M^{\rm
min}_{\rm c}<M_{\rm c, 1TP}$.\cite{k02,i04}

The TDU efficiency is defined as $\lambda=\frac{\Delta M_{\rm
dred}}{\Delta M_{\rm H}}$, where $ \Delta M_{\rm dred}$ is the mass
dredged up to the stellar surface during the thermal pulse, and
$\Delta M_{\rm H}$ is an increment of the core mass due to hydrogen
burning during the preceding interpulse period. $\lambda$ is a very
uncertain parameter. Karakas et al\cite{k02} showed a relation of
$\lambda$ to be
\begin{equation}
\lambda(N)=\lambda_{\rm max}[1-\exp(-N/N_{\rm r})], \label{eq:lamb}
\end{equation}
where $\lambda$ gradually increases towards an asymptotic
$\lambda_{\rm max}$ with the increase of $N$ (the progressive number
of thermal pulsation), and $N_{\rm r}$ is a constant determining how
fast $\lambda$ reaches $\lambda_{\rm max}$. In our work, $N_{\rm r}$
is taken as the value calculated from Eq.(49) in Ref.\cite{i04},
which can reproduce the results for $N_{\rm r}$ in Table 5 of
Ref.\cite{k02}. $\lambda_{\rm max}$ is given by (see Eq.(6) in
Ref.\cite{k02})
\begin{equation}
\lambda_{\rm max}=\frac{b_1+b_2M_0+b_3M^3_0}{1+b_4M_0^3},
\label{eq:lambmax}
\end{equation}
where coefficients $b_1$, $b_2$, $b_3$ and $b_4$ are shown in Table
8 of Ref.\cite{k02}.

In this paper, we take $\lambda$ as the value calculated from
Eq.(\ref{eq:lamb}).

\subsubsection{Intershell abundances}
During the every thermal pulse, the dredged mass $\Delta M_{\rm
dred}$ incorporates into stellar envelope. According to the
nucleosynthesis calculations by Boothroyd and Sackmann,\cite{bs88}
Marigo and Girardi\cite{mg07} gave the fitting of the abundances of
$^4$He,
$^{12}$C and $^{16}$O in the intershell region which lies between helium burning shell and hydrogen burning shell:\\
for $\Delta M_{\rm c}\leq$0.025$M_\odot$,\\
$^{\rm inter}X(^{4}{\rm He})=0.95+400(\Delta M_{\rm c})^2-20.0\Delta M_{\rm c},$\\
$^{\rm inter}X(^{12}{\rm C})=0.03-352(\Delta M_{\rm c})^2+17.6\Delta M_{\rm c},$\\
$^{\rm inter}X(^{16}{\rm O})=-32(\Delta M_{\rm c})^2+1.6\Delta M_{\rm c},$\\
for $\Delta M_{\rm c}>$0.025$M_\odot$,\\
$^{\rm inter}X(^{4}{\rm He})=0.70+0.65(\Delta M_{\rm c}-0.025),$\\
$^{\rm inter}X(^{12}{\rm C})=0.25-0.65(\Delta M_{\rm c}-0.025),$\\
$^{\rm inter}X(^{16}{\rm O})=0.02-0.065(\Delta M_{\rm c}-0.025).$\\
All the other isotopes are set to be zero in the intershell region
and the abundances are renormalized such that their sum is 1.0.
\subsubsection{The third dredge up and the hot bottom burning}
\label{sec:hbb} If the hydrogen envelope of an AGB star is
sufficiently massive, the hydrogen burning shell can extend to the
bottom of the convective region. This process is named the hot
bottom burning (HBB). For the HBB, we use a treatment similar to
that in Ref.\cite{gj93}. According to the model by Iben and
Renzini,\cite{ir83} Groenewegen and de Jong\cite{gj93} gave the most
suitable parameters for the fraction of newly dredged up matter
exposed to high temperatures at the bottom of the envelope $f_{\rm
HBB}=0.94$, the fraction of the envelope matter mixed down with the
bottom of the envelope $f_{\rm bur}=3\times10^{-4}$ and the exposure
time of the matter in the region of the HBB $t_{\rm
HBB}=0.0014\tau_{\rm ip}$. The temperature at the bottom of
convective envelope $T_{\rm bce}$ is given by Eq.(37) of
Ref.\cite{i04}.

For the drudged up masses, the quantities of materials added to the
envelope are:
\begin{equation}
\begin{array}{ll}
\Delta^4{\rm He}=&^{\rm inter}X(^{4}{\rm He})\Delta M_{\rm dred},\\
\Delta^{12}{\rm C}=&[(1-f_{\rm HBB})^{\rm
inter}X(^{12}{\rm C})\\
&+\frac{f_{\rm HBB}}{t_{\rm HBB}}
\int^{t_{\rm HBB}}_{0}X^{\rm HBB}_{12}(t){\rm d}t]\Delta M_{\rm dred},\\
\Delta^{13}{\rm C}=&[\frac{f_{\rm HBB}}{t_{\rm HBB}}
\int^{t_{\rm HBB}}_{0}X^{\rm HBB}_{13}(t){\rm d}t]\Delta M_{\rm dred},\\
\Delta^{14}{\rm N}=&[\frac{f_{\rm HBB}}{t_{\rm HBB}}
\int^{t_{\rm HBB}}_{0}X^{\rm HBB}_{14}(t){\rm d}t]\Delta M_{\rm dred},\\
\Delta^{16}{\rm O}=&[(1-f_{\rm HBB})^{\rm
inter}X(^{16}{\rm O})\\
&+\frac{f_{\rm HBB}}{t_{\rm HBB}} \int^{t_{\rm HBB}}_{0}X^{\rm
HBB}_{16}(t){\rm
d}t]\Delta M_{\rm dred},\\
\end{array}
\label{eq:hbb}
\end{equation}
where $X^{\rm HBB}(t)$ is the chemical abundance of the material
undergoing the HBB, and it is calculated in the way of Clayton's CNO
bicycle.\cite{c83} The details of Clayton's CNO bicycle can be seen
in Ref.\cite{c83} (also see Refs.\cite{i04,gj93}). The CNO bicycle
can be splited into the CN cycle and the ON cycle. The timescales in
ON cycle are many thousands of times longer than those required to
bring the CN cycle into equilibrium. Even in the most massive AGB
stars undergoing vigorous HBB, the ON cycle never approaches
equilibrium. Therefore, the effects of HBB mainly turn $^{12}$C into
$^{14}$N and the abundance of $^{16}$O is not changed much. In the
calculation of Clayton's CNO bicycle, the density on the base of the
convective envelope is given by Eq.(42) of Ref.\cite{i04} and the
analytic expressions for the nuclear reaction rates in Clayton's CNO
bicycle are cited from Ref.\cite{cf88}. The initial conditions of
$X^{\rm HBB}(t)$ are $X^{\rm HBB}(t=0)=\ ^{\rm inter}X$ for
$^{12}{\rm C}$ and $^{16}{\rm O}$, while $X^{\rm HBB}(t=0)=0$ for
$^{13}{\rm C}$ and $^{14}{\rm N}$.

After every thermal pulse, the chemical abundances of stellar
envelope $X^{\rm new}$s each are
\begin{equation}
\begin{array}{ll}
 X^{\rm new}&  \\
 =\frac{X^{\rm old}M_{\rm env}(1-f_{\rm bur})+\Delta
X+\frac{f_{\rm bur}M_{\rm env}}{t_{\rm HBB}}\int_0^{t_{\rm
HBB}}X(t){\rm d}t} {M_{\rm env}+\Delta M_{\rm dred}},& \\
\end{array}
\end{equation}
where the values of $\Delta X$ are given by expression
(\ref{eq:hbb}) and the values of initial conditions of $X(t)$ are
the values of $X(t=0)$ that are equal to $X^{\rm old}{\rm s}$.

In this work, we care about the ratio of the number of carbon atoms
to that of the oxygen atoms on the stellar surface and denote it as
$C/O$.

\subsubsection{Mass loss} \label{sec:ml} The mass loss rate of the
cool giant during AGB phase has a great influence on the chemical
evolution of the stellar surface. In our work, two
laws of the mass loss rate are considered: \\
(i) A mass loss relation suggested by Vassiliadis and
wood\cite{vw93} based on the observations, and it is given as
\begin{equation}
\begin{array}{ll}
\log \dot{M}= &-11.4+0.0123(P\\
                     &-100\max(M/M_\odot-2.5, 0.0)),\\
\end{array}
\label{eq:vwml}
\end{equation}
where $P$ is the Mira pulsation period in days, given by
\begin{equation}
\begin{array}{ll}
\log P = &-2.07+1.94 \log(R/R_\odot)\\
 &-0.90\log(M/M_\odot).\\
 \end{array}
\end{equation}
When $P \geq 500$ days, the steady super-wind phase is modeled by
the law
\begin{equation}
\dot{M}(M_\odot {\rm yr}^{-1})=2.06\times
10^{-8}\frac{L/L_\odot}{v_\infty},
\end{equation}
where $v_\infty$ is the terminal speed of the super-wind in km
s$^{-1}$; we use $v_\infty$=15 km s$^{-1}$ in this paper.\\
(ii) A mass loss rate similar to Reimers' formula, which is given by
Bl\"{o}cker\cite{b95} according to the simulations of shock-driven
winds in the atmospheres of Mira-like stars\cite{b88} and expressed
as
\begin{equation}
\dot{M}=4.83\times10^{-9}M^{-2.1}L^{2.7}\dot{M}_{\rm Reimers},
\label{eq:bml}
\end{equation}
where $\dot{M}_{\rm Reimers}$ is given by expression (\ref{eq:rml}),
and $\eta=0.02$.\cite{sj07}

In the other stellar evolutionary phases, the mass loss rates are
given by Reimers' formula\cite{r75}
\begin{equation}
\dot{M}=-4.0\times10^{-13}\eta\frac{LR}{M}{\rm M_\odot yr^{-1}},
\label{eq:rml}
\end{equation}
where $L$, $R$, $M$ are the stellar luminosity, radius and mass in
solar units, respectively, and the free parameter is $\eta$=0.5.

\section{Results}
We construct a set of models with different input parameters. Table
1 lists all the cases considered in the present work. We calculate
$1\times 10^5$ single stars for each case.
\begin{table*}
\caption  {Parameters of the models. $\theta$ means the coefficient
of oxygen overabundance.}
\begin{tabular}{|c|c|c|}
\hline
case & coefficient $\theta$& Mass loss rate\\
\hline
case 1& 1.0&Eq.(\ref{eq:vwml})\\
case 2& 2.0&Eq.(\ref{eq:vwml})\\
case 3& 3.0&Eq.(\ref{eq:vwml})\\
case 4& 5.0&Eq.(\ref{eq:vwml})\\
case 5& 2.0&Eq.(\ref{eq:bml})\\
case 6& 5.0&Eq.(\ref{eq:bml})\\
\hline
\end{tabular}
\end{table*}
\subsection{The Effects of Parameters}
\textbf{Positions in Hertzsprung-Russel (HR) -diagram} In Fig.1, the
different evolutionary phases during TP-AGB in HR-diagram are shown.
For the same mass loss rate (see Figs.1(a)-1(d), or Figs.1(e) and
1(f)), it is more difficult to form carbon stars with the increased
of oxygen overabundance coefficient $\theta$ because the stars must
drudge up more $^{12}$C to the stellar surface in order to obtain
$C/O>1$. When $\theta=5.0$, the initial oxygen abundance is too high
so that the dredged up $^{12}$C is not enough for $C/O>1$, thus
there is no carbon star (see Fig.1(d)). On the other hand, for the
same coefficient $\theta$ (Figs.1(b) and 1(e), or Figs.1(d) and
1(f)), the mass loss rate taken as the value calculated from
expression (\ref{eq:vwml}) is more favorable for preventing the
carbon stars for forming than that described by Eq.(\ref{eq:bml}).
As we can see from Fig.1, the stars with the initial masses from
2.0$M_\odot$ to 4.0$M_\odot$ may become the carbon stars.


\begin{figure}
\includegraphics[totalheight=3.2in,width=3.8in,angle=-90]{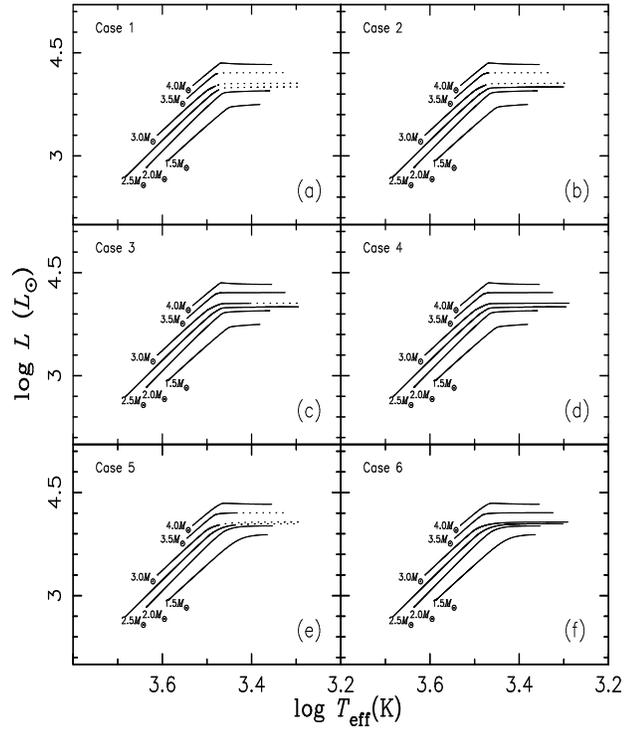}
\caption{ The positions of TP-AGB stars in HR-diagram for the
different cases. The solid lines mean O-rich TP-AGB stars and the
dotted lines represent carbon stars. Stellar initial masses are
shown by the figures in the side of the lines.}
\end{figure}

\textbf{Variety of $C/O$} Figure 2 shows the varieties of $C/O$
during the TP-AGB phase. For a star with an initial mass
1.5$M_\odot$, $C/O$ is almost a constant as shown in Fig.2(a).
Figures 2(b)-2(f) show, for the stars with initial masses ranging
from 2.0$M_\odot$ to 4.0$M_\odot$, all of $C/O$s are changed but
their varieties are different form each other. For a star with an
initial mass 2.0$M_\odot$ (see Fig.2(b)), the core mass $M_{\rm c}$
is lower than $M_{\rm c}^{\rm min}$ when it just evolves to the
TP-AGB. It takes a long time to satisfy a certain core mass $M_{\rm
c}^{\rm min}$ of the TDU. Therefore, it is hard to reach $C/O>1$.
There is no carbon star with an initial mass 4.0$M_\odot$ (see
Fig.2(f)). The main reason is that the star with an initial mass
higher than 4.0$M_\odot$ has a convective envelope which is thick
enough, thus the temperature at the bottom of the convective
envelope, $T_{\rm bce}$, is high enough to undergo the HBB. The HBB
turns $^{12}$C in the drudge up materials into $^{14}$N and prevents
the carbon stars for forming.\cite{i04} From Fig.2(e), we can see
obviously the largest variety of $C/O$ in case 1 and the smallest in
case 4 for the stars with the same initial masses. It means that
case 4 is the most unfavorable case for forming the carbon star in
our model.

\begin{figure}
\includegraphics[totalheight=3in,width=3.5in,angle=-90]{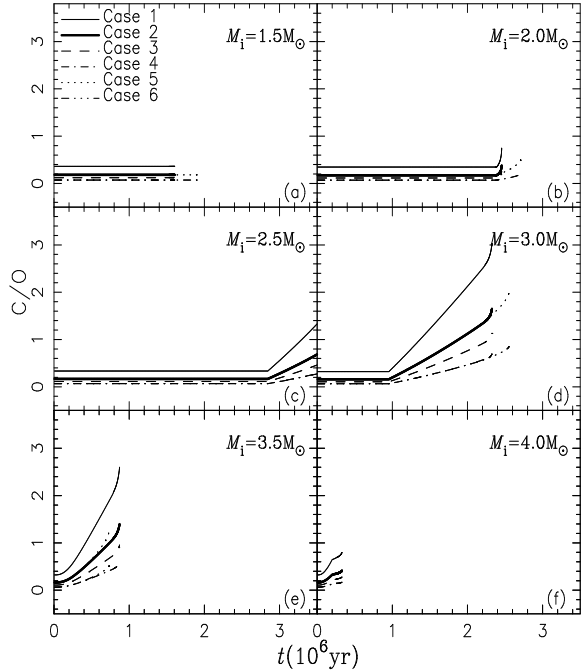}
\caption{The varieties of $C/O$ during the TP-AGB phase in the
different cases and the initial masses. Stellar initial masses are
shown in right top corners. The cases the different styles of lines
correspond to are shown in the left top region of Fig.2(a).}
\end{figure}

\textbf{Lifetimes} Figure 3 gives the stellar lifetimes of the
different evolutionary phases. The phase of the carbon stars is a
small fraction of the whole TP-AGB. From Fig.3, we see that the
peaks of lifetimes for the carbon stars each appear at 3.1$M_\odot$,
which are different from the peaks at 2.5$M_\odot$ for the whole
TP-AGB stars. The main reason is as follows: for a star with an
initial mass 3$M_\odot$, $M_{\rm c}^{\rm min}\simeq M_{\rm{c,1TP}}$
when it evolves into the TP-AGB phase. After a few thermal pulses
the star begins to undergo the TDU. At the same time, according to
expression (\ref{eq:lambmax}), the $\lambda_{\rm max}$ is about 0.8
for a 3$M_\odot$ star (also see Fig.4 of Ref.\cite{k02}). Therefore,
the TDU has a high efficiency $\lambda$. The increases of its core
mass and the mass loss of the envelope are slow so that the star
takes a longer time to evolve to the end of TP-AGB.

Figure 3 shows that the oxygen overabundance coefficient $\theta$
has a significant influence on the lifetimes of the carbon stars.
With the increase of coefficient $\theta$, the lifetimes of the
carbon stars decrease. There is no carbon star when  $\theta$
increases up to 5.0 (see Fig.3(d)). In addition, Fig.3 indicates
that the mass loss rate also has an influence on the lifetimes of
the carbon stars. For the same values of coefficient $\theta$
(Figs.3(b) and 3(e)), the lifetimes of the carbon stars with the
mass loss rate as expressed by expression (\ref{eq:vwml}) are
shorter than that given by expression (\ref{eq:bml}).

\begin{figure}
\includegraphics[totalheight=3.2in,width=3.8in,angle=-90]{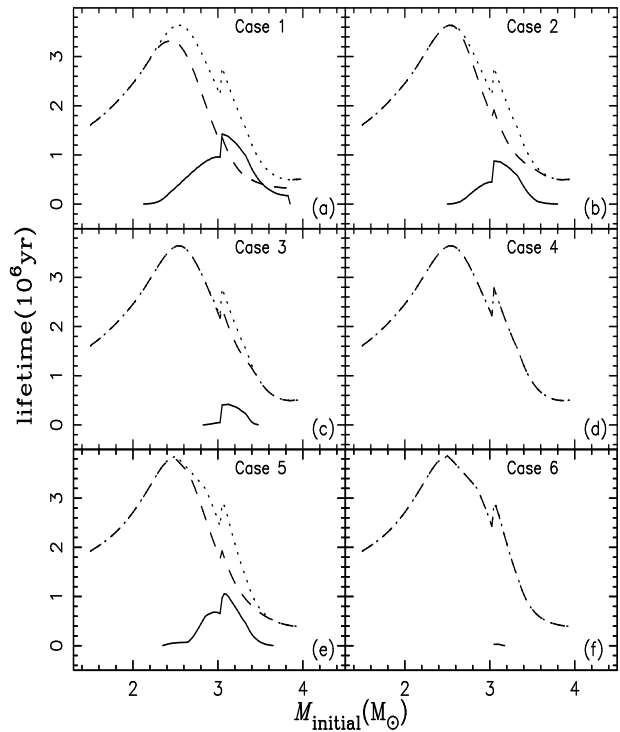}
\caption{The stellar lifetimes of the different evolutionary phases.
The dot-dashed, dashed and solid lines represent the whole TP-AGB
stars, O-rich stars and Carbon stars, respectively.}
\end{figure}

\textbf{Mass distribution} Figure 4 shows the distribution of
numbers (all the numbers are normalized to 1) of the modeled TP-AGB
stars in the Galactic bulge as a function of initial stellar mass.
Figure 4 shows that the initial masses of the progenitors of the
carbon stars range mainly from 2.5$M_\odot$ to 3.5$M_\odot$ and the
peaks of the initial masses of the carbon stars each appear at
3.1$M_\odot$, which is in good agreement with the case in Fig.3.
Comparing the cases in Fig.4, we easily see the effects of the
parameters $\theta$ and the mass loss rate. The results are
consistent with the descriptions in Fig.3.

\begin{figure}
\includegraphics[totalheight=3in,width=3.8in,angle=-90]{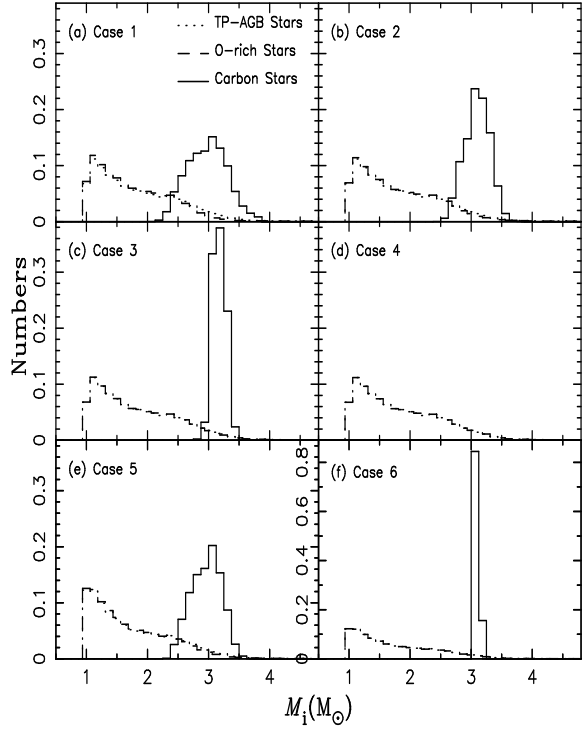}
\caption {The distributions of numbers (all the numbers are
normalized to 1) of the modeled TP-AGB stars in the Galactic bulge
as a function of initial stellar mass. The cases the different
styles of lines correspond to are shown in the right top region of
Fig.4(a).}
\end{figure}

\subsection{Lifetimes and the Numbers of Carbon Stars}
In this subsection, we discuss the rough properties of the modeled
population of carbon stars in the Galactic bulge and then proceed to
make a more detailed comparison among the influences from the
different assumptions.
\begin{table*}
\centering
\begin{minipage}{140mm}
\caption {The different models of TP-AGB population. The first
column gives case number according to Table 1.
 Columns 2, 4 and 6 give the numbers of TP-AGB stars, O-rich stars and carbon stars in the bulge, respectively.
 Columns 3, 5 and 7 show the average lifetimes of TP-AGB, O-rich and carbon stars, respectively.
 Column 8 is the ratios of the number of O-rich stars to that of carbon stars.}
\begin{tabular*}{200mm}{|c|c|c|c|c|c|c|c|}
\cline{1-8} \multicolumn{1}{|c|}{}&\multicolumn{2}{|c|}{TP-AGB
Stars}&\multicolumn{2}{|c|}{O-rich
Stars}&\multicolumn{2}{|c|}{Carbon
Stars}&\multicolumn{1}{|c|}{The ratios}\\
\cline{2-7}
Cases& Number&Lifetime&Number&Lifetime&Number&Lifetime&of\\
  & $10^5$&$10^6$years&$10^5$&$10^6$years&$10^5$&$10^6$years&the number\\
\cline{1-8}
1&2&3&4&5&6&7&8\\
case 1& 113.23&1.46&107.03&1.38&6.20&0.48&17.3\\
case 2& 113.23&1.46&110.94&1.43&2.28&0.29&48.7\\
case 3& 113.23&1.46&112.59&1.45&0.63&0.17&178.7\\
case 4& 113.23&1.46&113.23&1.46&0&0&$\infty$\\
case 5& 138.52&1.79&135.07&1.74&3.45&0.37&39.2\\
case 6& 138.52&1.79&138.51&1.79&0.02&0.02&6925.5\\
\cline{1-8}
\end{tabular*}
\end{minipage}
\end{table*}

Table 2 gives the different models of TP-AGB population in the
Galactic bulge.

(i) The mass loss rate is taken as the value calculated from
expression (\ref{eq:vwml}) (cases 1, 2, 3 and 4): the numbers of the
carbon stars in the bulge may range from 0 (case 4) to $6.02\times
10^5$ (case 1). Their average lifetimes range from 0 (case 4) to
$0.48\times 10^6$ years (case1). The ratios of the number of O-rich
stars to that of the carbon stars may range from 17.3 (case 1) to
$\infty$ (case 4), where $\infty$ means that there is no carbon star
in this case. The above results suggest that the oxygen
overabundance has great effects on the numbers and the lifetimes of
the carbon stars in the bulge. As shown in Table 2, the numbers and
the lifetimes of the carbon stars decrease with coefficient $\theta$
increasing. It is worth mentioning that Rich and Origlia\cite{ro05}
indicated that in the bulge M giants oxygen overabundance is twice
as big as that in the solar neighborhood. According to our model,
$\theta=2.0$, which is insufficient for the lack of the carbon stars
in the Galactic bulge. The carbon stars cannot be formed when
$\theta=5.0$ in our model, but the initial oxygen abundance is much
higher than the observation.

(ii) The mass loss rate is taken as the value calculated from
expression (\ref{eq:bml}) (cases 5 and 6): Table 2 shows that the
numbers and the average lifetimes of the carbon stars may range from
$0.02\times 10^5$ (case 6) to $3.45\times 10^5$ (case 5) and from
$0.02\times 10^6$ years (case 6) to $0.37\times 10^6$ years (case
5), respectively. It is obvious that the oxygen overabundance is
favorable for preventing the carbon stars from forming, however, it
is not enough to explain the lack of carbon stars in the bulge
unless the oxygen overabundance coefficient $\theta$ is equal to
5.0. This is consistent with the above result.

(iii) For the same coefficient $\theta$ (cases 2 and 5, or cases 4
and 6), we can see from Table 2 that the mass loss rate taken as the
value calculated from expression (\ref{eq:vwml}) is helpful for
explaining the lack of the carbon stars in the Galactic bulge.
Recently, Willson\cite{w07} gave a detailed description on ${\rm
d}(\log \dot{M})/{\rm d}(\log L)$ (see Fig.1 of Ref.\cite{w07}). The
average mass loss rate calculated from expression (\ref{eq:vwml}) is
higher than that calculated from expression (\ref{eq:bml}) during
the TP-AGB in our simulations. The higher the mass loss rate is, the
more quickly the envelope mass $M_{\rm env}$ will decrease.
According to expression (\ref{eq:tip}), the interpulse period
$\tau_{\rm ip}$ increases with envelope mass decreasing. A long
$\tau_{\rm ip}$ can reduce the TDU progressive number and TDU
efficiency $\lambda$. The large mass loss rate is unfavorable for
forming the carbon stars. The environment in the Galactic bulge is
significantly different from that in the solar neighborhood, for
example, the higher irradiating luminosity and the stellar density.
These factors may enhance the stellar mass loss rate. We suggest
that the mass loss rate may be a factor of controlling the ratio of
the number of O-rich stars to the number of carbon stars in the
Galactic bulge.


\section{Conclusion}
Employing the population synthesis, we make a detailed study of the
TP-AGB stars in the Galactic bulge. We emphasize the relationship
between the formation of the carbon stars and the relevant
parameters. The effects of the oxygen overabundance coefficient and
the mass loss rate on the ratio of the number of the carbon stars to
the that of O-rich stars are discussed. We find that the ratio of
the number of the carbon stars to that of O-rich stars is greatly
affected by the oxygen overabundance coefficient $\theta$. However,
comparing with the present observations of  the oxygen overabundance
in the Galactic bulge, we believe that the oxygen overabundance
which is about twice as large as that in the solar neighborhood is
insufficient to explain the lack of the carbon stars in the bulge.
In addition, we obtain that the mass loss rate also is a controlling
factor in the ratio of the number of carbon stars to that of O-rich
stars. The larger mass loss rate is helpful for preventing the
carbon stars from forming. There may be a more larger mass loss rate
in the Galactic bulge than in the solar neighbourhood. Further work
on this is desirable.

\section*{Acknowledgment}
We are grateful to Dr. Izzard for providing his Doctor's thesis.
L$\ddot{u}$ Guo-Liang thanks Prof. Zhanwen Han for helpful
discussion and offering CPU time. We thank Dr. Hoernisa Iminniyaz
for correcting the English language in the manuscript.


\begin{thebibliography}{99}
\bibitem{k02}Karakas A I, Lattanzio J C and Pols O R 2002 {\it Publ. Astron. Soc. Aust.}
\textbf{19} 515
\bibitem{v93}Vassiliads E and Wood P R 1993 {\it Astrophys. J.} \textbf{413} 641
\bibitem{b91}Bowen G H and Willson L A 1991 {\it Astrophys. J.} \textbf{375} L53
\bibitem{i83}Iben I and Renzini A 1983 {\it Ann. Rev. Astr. Ap.} \textbf{21} 271
\bibitem{w93}Whitelock P A 1993 {\it IAUS} \textbf{153} 39
\bibitem{t91}Tyson N D and Rich R M 1991 {\it Astrophys. J.} \textbf{367} 547
\bibitem{f06}Feast M W 2006 {\it ASP Conference Series} (in press) astro-ph/0609318
\bibitem{h95}Han Z, Eggleton P, Podsiadlowski P and Tout C A 1995 {\it Mon. Not. R. Astron. Soc.} \textbf{277} 1443
\bibitem{z02}Zhang F H, Han Z, Li L F and Hurley J R 2002 {\it Chinese Physics Letters} \textbf{19} 1734
\bibitem{h03}Han Z, Podsiadlowski P, Maxted P F L and Marsh T R 2003 {\it Mon. Not. R. Astron. Soc.} \textbf{341} 669
\bibitem{ktg93}Kroupa P, Tout C A and Gilmore G 1993 {\it Mon. Not. R. Astron. Soc.} \textbf{262} 545
\bibitem{a97}Andrew McWilliam 1997 {\it Ann. Rev. Astr. Ap.} \textbf{35} 503
\bibitem{g01}Gilmore G 2001 {\it ASP Conference Series} \textbf{230} 3
\bibitem{z05}Zhou W F, Wu Y F, Wei Y and Ju B G 2005 {\it Chin. Phys.} \textbf{14} 863
\bibitem{h00}Hurley J R, Pols O R and Tout C A 2000 {\it Mon. Not. R. Astron. Soc.} \textbf{315} 543
\bibitem{i04}Izzard R G, Tout C A, Karakas A I and Pols O R 2004 {\it Mon. Not. R. Astron. Soc.} \textbf{350} 407
\bibitem{I04}Izzard R G 2004 {\it Doctor thesis} University of Cambridge
\bibitem{ro05}Rich R M and Origlia L 2005 {\it Astrophys. J.} \textbf{634} 1293
\bibitem{AG89}Anders E and Grevesse N 1989 {\it Grochim. Cosmochim. Acta} \textbf{53} 197
\bibitem{sg00}Salasnich B, Girardi L, Weiss A and Chiosi C 2000 {\it Astron. Astrophys.} \textbf{361} 1023
\bibitem{gj93}Groenewegen M A T and de Jong T 1993 {\it Astron. Astrophys.} \textbf{267} 410
\bibitem{mg07}Marigo P and Girardi L 2007 {\it Astron. Astrophys.} \textbf{469} 239
\bibitem{bs88}Boothroyd A I and Sackmann I J 1988 {\it Astrophys. J.} \textbf{328} 653
\bibitem{ir83}Iben I and Renzini A 1983 {\it Annual Review of Astronomy and Astrophysics} \textbf{21} 271
\bibitem{c83}Clayton D D 1983 {\it Principles of Stellar Evolution and Nucleosynthesis}
Chicago p.390-410
\bibitem{cf88}Caughlan G R and Fowler W A 1988 {\it Atomic Data and Nuclear Data Tables} \textbf{40} 283
\bibitem{vw93}Vassiliadis E and Wood P R 1993 {\it Astrophys. J.} \textbf{413} 641
\bibitem{b88}Bedijn P J 1988 {\it Astron. Astrophys.} \textbf{205} 105
\bibitem{b95}Bl\"{o}cker T 1995 {\it Astron. Astrophys.} \textbf{297} 727
\bibitem{sj07}Stancliffe R J and Jeffery C S 2007 {\it Mon. Not. R. Astron. Soc.} \textbf{375} 1280
\bibitem{r75}Reimers D 1975 {\it Mem. Soc. R. Sci. Liege} \textbf{8} 369
\bibitem{w07}Willson L A 2007 astro-ph/07043589
\end{thebibliography}
\end{document}